\documentstyle[prl,preprint,aps,epsf]{revtex}
\begin{document}

\title{Bifurcation of Periodic Instanton in Decay-Rate Transition}

\author{ Hyun-Soo Min$^a$, Hungsoo Kim$^b$, D. K. Park$^a$, Soo-Young Lee$^a$,
Sahng-Kyoon Yoo$^c$, and Dal-Ho Yoon$^d$}

\address{$^a$ Department of Physics, Kyungnam University, Masan, 631-701,
Korea \\ 
$^b$ Department of Physics, Korea Advanced Institute of Science and Technology,
Taejon, 305-701, Korea \\
$^c$ Department of Physics and Basic Science Research Institute, Seonam
University, Namwon, 590-711, Korea \\
$^d$ Department of Physics, Chongju University, Chongju, 360-764, Korea}
\date{\today}
\maketitle

\begin{abstract}
We investigate a bifurcation of periodic instanton in Euclidean 
action-temperature diagram in quantum mechanical models.
It is analytically shown 
that multiple zero modes of fluctuation
operator should be arised at bifurcation points. 
This fact is used to derive a condition for the appearance 
of bifurcation points in action-temperature diagram.
This condition enables one to compute the number of bifurcation
points for a given quantum mechanical system and hence, to understand
the whole behaviour of decay rate. It is explicitly shown that the 
previous criterion derived by nonlinear perturbation or 
negative-mode consideration is special limit of our case.
\end{abstract}

\newpage
\section{Introduction}

Recently, much attention is paid to quantum-classical  
decay-rate transition in 
various branches of physics such as spin tunneling systems\cite{ch97,li98,pa99},
field theoretical models\cite{ha96,ku97,dpark99}, 
and cosmology\cite{lin83,fer95,kim99}. An important 
issue in this field is to determine the type of the decay-rate transition. 
At zero temperature the decay rate by quantum tunneling is governed 
by instanton or bounce\cite{col77}, and at intermediate
range of temperature  periodic 
instanton\cite{kh91}
plays an important role. At high temperatures the decay 
takes place mainly by thermal activation, which is represented 
by sphaleron\cite{man83}. Usually the decay-rate transition means 
the transition between quantum tunneling regime dominated by 
periodic instanton and thermal activity regime dominated by sphaleron.    

The type of the decay-rate transition 
is completely determined by Euclidean action-temperature 
diagram as shown in Ref.\cite{ch92}.  
The typical action-temperature diagrams which appear frequently 
in quantum mechanical or field theoretical models 
can be classified by three types(see Fig. 1).  
Fig. 1(a) represents 
a smooth second-order transition in which there is no bifurcation 
of periodic instanton and hence, the decay rate varies smoothly 
with increasing 
temperature. In Fig. 1(b) there is one bifurcation point and the decay 
rate exhibits an abrupt change at the crossover from periodic 
instanton to sphaleron. Thus this is a typical diagram for 
the first-order transition. 
Fig. 1(c) has two bifurcation points and the decay-rate transition 
between regimes dominated by periodic instanton and sphaleron, respectively,
is second order. However, 
there exists  
a sharp change 
of decay rate in quantum tunneling regime in this case.
For         
convenience we call the types of decay-rate transition associated to  
Fig. 1(a), Fig. 1(b), and Fig. 1(c) 
type I, type II, and type III, respectively.

In order to determine the type of transition completely from 
action-temperature diagram one has to compute a periodic instanton
explicitly. However, the explicit derivation of periodic instanton
in field theoretical models is very complicated and sometimes 
impossible. Hence, it is important to develop a method which determines
the type of decay-rate transition without computing the periodic
instanton in the full range of temperature.

The researches along this direction were done recently by using 
nonlinear perturbation method\cite{go97,mu99} or counting the number
of negative modes of full hessian operator around sphaleron\cite{lee99}.
Although these two methods start from completely different point
of view, both method derive a same criterion for the sharp
first-order transition.
Since, however, these two methods use only periodic instanton near
sphaleron, it is impossible to distinguish type III transition from
type I by these methods.

The purpose of this letter is to develop a more powerful method which
not only distinguish type III from type I but also enable one to 
understand the whole behaviour of decay-rate transition by exploring
the properties of bifurcation point. In Sec. II we will show analytically
that multiple zero modes should be arised at bifurcation point. This fact
is used to derive a criterion for the appearance of bifurcation
point in action-temperature diagram, which will be explored in Sec. III.
Also, application of this criterion to simple quantum mechanical model
will be presented in the same section. 
It is shown that the criterion derived by nonlinear perturbation method is 
only special limit of our result.
In the final section a brief 
conclusion is given.


\section{Zero modes at bifurcation point}

The decay rate at finite temperatures can be evaluated by Euclidean 
action 
of periodic instanton as
\begin{equation}
\Gamma (T) = A e^{-S_E [x]}
\end{equation}
where $A$ is pre-exponential factor.  
The Euclidean action $S_E[x]$ is represented as
\begin{equation}
S_E[x] = \int _{0}^{\beta} d\tau [ \frac{1}{2}\dot{x}^2 + V(x)],
\end{equation}
where $\beta$, period at Euclidean space, is
given by inverse of temperature, and 
we assume, for convenience, particle mass is unity.
The equation of motion for periodic instantion is
  \begin{equation}
  \ddot{x} = V'(x),
  \end{equation}
where the prime denotes the coordinate 
derivative, i.e, $V'(x) =d V/dx$, and
the Euclidean energy $E$ is  
\begin{equation}  
E = V(x) - \frac{1}{2} \dot{x}^2.  \label{energy}  
\end{equation}

Now, let us consider periodic instantons $x_0(\tau)$ with period $\beta_0$ and 
$x_1(\tau)$ with $\beta_1 = \beta_0 + \delta \beta$. 
Then we can write  
\begin{eqnarray}  
x_1(\tau) &=& x_0(\frac{\beta_0}{\beta_1} \tau ) +   
            \frac{\delta \beta}{\beta_0} \eta(\tau)  \nonumber \\
  &\simeq& x_0 (\tau) - \frac{\delta \beta}{\beta} 
  [ \tau \dot{x_0} -\eta(\tau)]  \label{x1}  
\end{eqnarray}
where $\eta(\tau)$ is some periodic function which can be determined from the
fact that $x_0(\tau)$ and $x_1(\tau)$ must be a solutions of equation of 
motion. Hence, $\eta(\tau)$ has to satisfy
\begin{equation}  
\hat{M} \eta = -2\ddot{x_0},  \label{eqmoeta}  
\end{equation}
where $\hat{M}$ is the fluctuation operator at $x_0$;  
\begin{equation}  
\hat{M} = -\frac{d^2}{d\tau^2} + V''(x_0).  
\end{equation}
Using Eqs. (\ref{energy}) and (\ref{x1}), one can obtain the ratio 
between energy and period differences;  
\begin{equation}  
\frac{\delta E}{\delta \beta} = 
\frac{1}{\beta}   [V'(x_0) \eta +  \dot{x_0}^2 -  \dot{x_0}\dot{\eta}].  
\label{newla}
\end{equation}
It is worthwhile noting that the right-hand side of Eq.(\ref{newla}) is a 
constant of motion.
Since bifurcation point takes place when $\frac{d\beta}{d E} =0$,  
the absolute value of the right-hand side of the above equation 
must 
diverge at this point. 
This means that $\eta$ must be a singular function at the same point.
In order for $\eta$ to be singular we need another zero mode different from 
well-known one, i.e., $\dot{x_0}$ originated from time translational symmetry.
This fact can be easily shown as follows.
If we expand the right-hand side of  Eq.(\ref{eqmoeta}) in terms of 
eigenstates $|\xi_n>$ of $\hat{M}$,
the equation for $\eta$ becomes
\begin{equation}
   \hat{M} |\eta > = \sum^{ }_n {}^{\prime} a_n |\xi_n>,
\end{equation}
where $a_n$'s are expansion coefficients and the prime denotes 
that the sum excludes the known zero mode $|\dot{x_0}>$ due to its
orthogonality to $|\ddot{x_0}>$.
Then $|\eta>$ can be simply written as
\begin{equation}
   |\eta> = \sum^{ }_n  {}^{\prime} \frac{a_n}{h_n} |\xi_n>,
\end{equation}
where $h_n$'s are corresponding eigenvalues of $\hat{M}$.
Here, it is shown that in order to get infinite magnitude of $|\eta>$
at bifurcation point, at least one of $h_n$'s should be zero.This zero 
mode 
does not correspond to the known zero mode($\dot{x_0}$) but new zero mode at
bifurcation point.
This fact is numerically explored at Ref.\cite{lee99}.

In the next section we derive a condition for appearane of bifurcation
point in action-temperature diagram.

\section{Condition for bifurcation point}
As mentioned in the previous section, the zero mode satisfies
the following second-order ordinary differential equation;
\begin{equation}
   -\frac{d^2 y}{d\tau^2} + V''(x_0) y = 0.
\end{equation}
Since this equation is second order, there are two independent
solutions; one is the known one, $y_1=\dot{x_0}$, and another
is 
\begin{equation}
   y_2(\tau) = y_1(\tau) \int^{\tau} \frac{d\tau'}{y_1^2(\tau')}
       = \dot{x_0}(\tau) \int^{\tau} \frac{d\tau'}{\dot{x_0}^2(\tau')}.
\end{equation}
This fact does not mean that the periodic instanton $x_0$ always have
two independent zero modes, because in general  $y_2(\tau)$ does not satisfy
periodic boundary condition. 
In fact, $y_2(\tau)$ satisfies the physically relevant boundary condition
at only bifurcation point.
From the symmetry of $x_0(\tau)$ one can conjecture that
in order for $y_2(\tau)$ to be an another zero mode $\dot{x_0}(\tau)$ and
$y_2(\tau)$ must have common turning points. 
Using Eq.(\ref{energy}), this conjecture can be mathematically written as
\begin{equation}
   \int_{x_-}^{x_+} dx \frac{1}{(V(x) - E )^{3/2}} = 
  \frac{2}{V'(x_-)\sqrt{V(x_-) - E }} - \frac{2}{V'(x_+)\sqrt{V(x_+)-E}}
\end{equation}
where $x_+$ and $x_-$ are turning points.
This is not a desirable expression for the condition of bifurcation since
both sides of this equation are infinite. 
However, if one uses a relation
\begin{equation}
   \frac{d}{dx} ( \frac{1}{\sqrt{V(x) - E}} ) = 
    - \frac{V'(x)}{2 (V(x) - E)^{3/2}},
\end{equation}
the condition for the appearance of bifurcation becomes
\begin{eqnarray}
  f(E) &\equiv& V'(x_-) \int_{x_s}^{x_+} dx \frac{V'(x_+) - V'(x)}{(V(x) - E)^{3/2}}
  +V'(x_+) \int_{x_-}^{x_s} dx \frac{V'(x_-) - V'(x)}{(V(x) - E)^{3/2}}
  + \frac{2 [ V'(x_-) - V'(x_+)]}{\sqrt{V(x_s) - E}}   \nonumber \\
   & =& 0,
  \label{final}
\end{eqnarray}
where $x_s$ is sphaleron solution, i.e, position of barrier top.
It is easily shown that the values of $f(E)$ at minimum energy $E=E_{min}$
and at sphaleron energy $E=E_s$ are positive and zero, respectively, i.e.,
$f(E_{min}) > 0$ and $f(E_s) =0$.
Since at zeros of $f(E)$ bifurcations take place, we can determine 
the type of decay rate from the number of zeros of $f(E)$ in the 
range of $E_{min} < E < E_s$;  non-existence of zeros  means type I, one 
and two zeros correspond to type II and  type III, respectively.

Now, let us apply our criterion (\ref{final}) to simple quantum 
mechanical model whose potential is 
\begin{equation}    
V(x) = \frac{4+\alpha}{12}- \frac{1}{2}x^2 - \frac{\alpha}{4}x^4             
+\frac{\alpha+1}{6}x^6 - \frac{\gamma}{3} x^3.   
\label{qmpotential}
\end{equation}
Numerical calculation using the criterion (\ref{final}) yields
Fig. 2 which shows the type of decay rate along the potential 
parameter $\alpha$ and$\gamma$. 
The dashed line distinguishing type III from type I cannot be 
obtained by the criterion of Ref.\cite{mu99} since type I and III 
have same behavior in the vicinity of sphaleron. Now, let us derive the 
previous result of Ref.\cite{mu99} by imposing a simple restriction on $f(E)$.
Since the sufficient criterion for first-order transition has been derived by
nonlinear perturbation at sphaleron,  it can be obtained from a 
simple restriction on value of $f(E)$ near sphaleron;
\begin{equation}
     f(E_s -\epsilon) < 0,
\label{fe}
\end{equation}
where $\epsilon$ is a infinitesimal positive number.

Since one can put $x_s=0$ without loss of generality,
two integrals of Eq.(\ref{final}) can be properly expanded in terms
of following integrals;
\begin{eqnarray}
   G_n &\equiv& \int_{0}^{x_+} dx \frac{x^n}{\sqrt{x_+ - x}(x-x_-)^{3/2}}, \\
   H_n &\equiv& \int_{x_-}^{0} dx \frac{x^n}{\sqrt{x - x_-}(x_+ -x)^{3/2}}.
\end{eqnarray}
Since $G_n$ and $H_n$ can be exactly evaluated by $x_{\pm}$, we can
expand Eq.(\ref{fe}) as increasing power series of $x_{\pm}$.
Then we get first non-zero term as
\begin{equation}
    -\frac{5}{3} \frac{V'''(0)^2}{V''(0)} + V''''(0) < 0,
\end{equation}
where we leave out irrelevant factor.
This is the same with the result of nonlinear perturbation method, which means
that the previous result is a special case of our condition for bifurcation.

\section{Conclusion}

From the fact that an additional zero mode appears at bifurcation points,
we derive the condition Eq.(\ref{final}) for occurrence of bifurcation 
in action-temperature diagram. Using this condition, one can count the 
number of bifurcations in allowed energy range and hence,
understand the whole behaviour of decay-rate transition.  
The sufficient criterion for the first-order transition 
can be also derived by imposing a restriction near sphaleron, without 
use of nonlinear perturbation\cite{mu99} or negative-mode 
consideration\cite{lee99}. 
From the restriction it can be understood that 
the previous criterion is a condition to have odd number of bifurcations, 
so that with the criterion one cannot distinguish type III from type I.
We hope that the idea for bifurcation in this paper could be generalized 
to be applicable to 
field theoretical models, which, however, might be non-trivial.
Work on this direction is in progress.


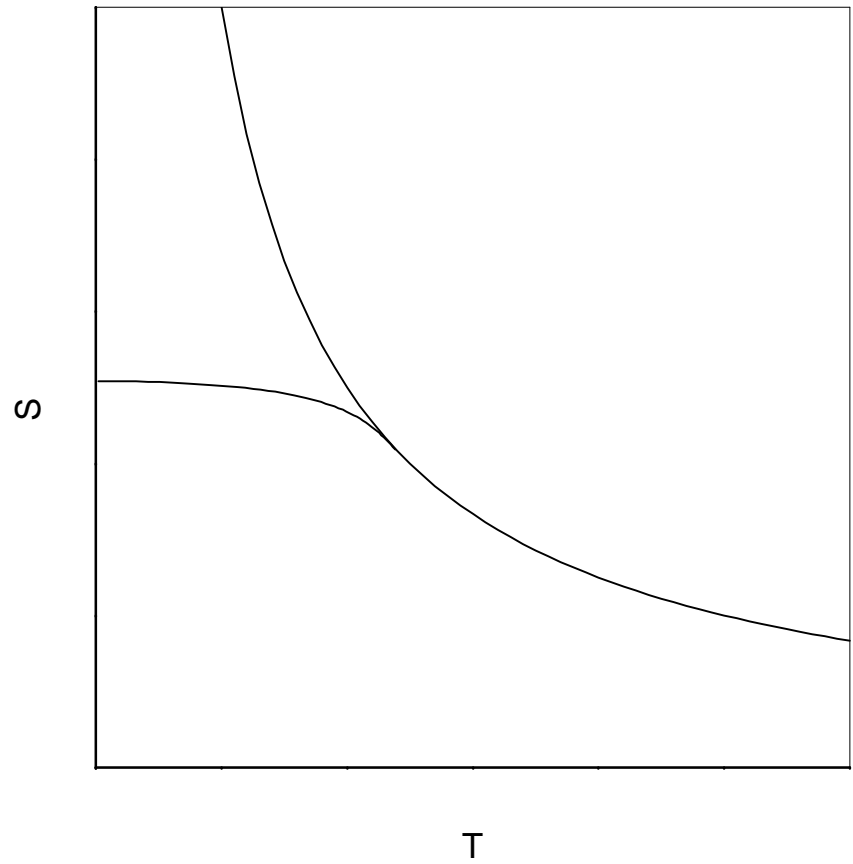
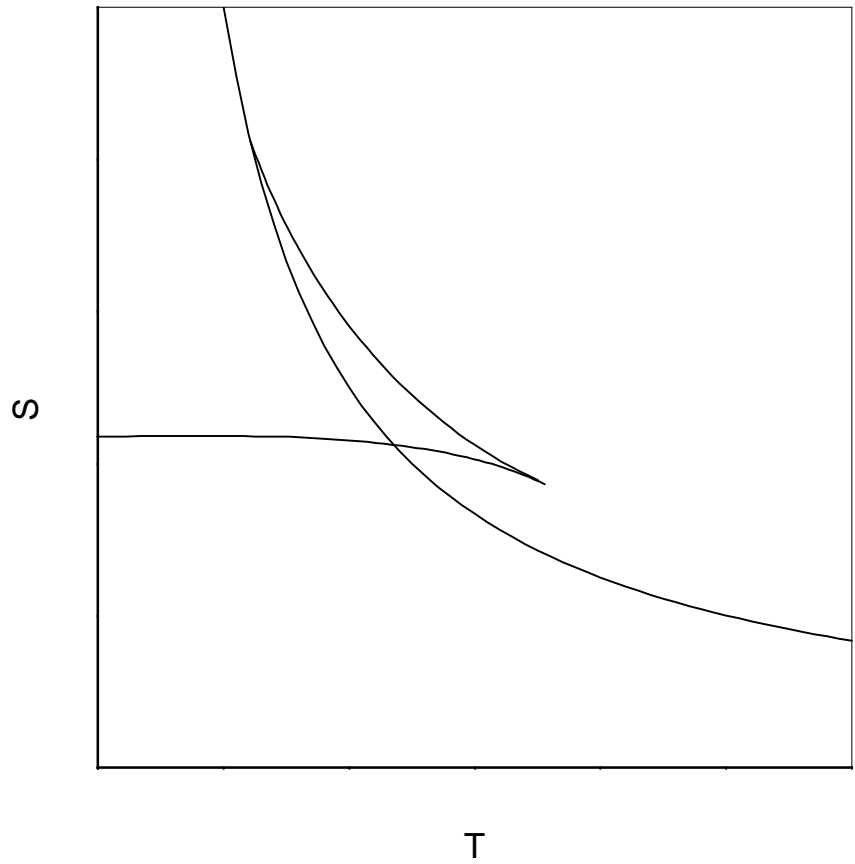
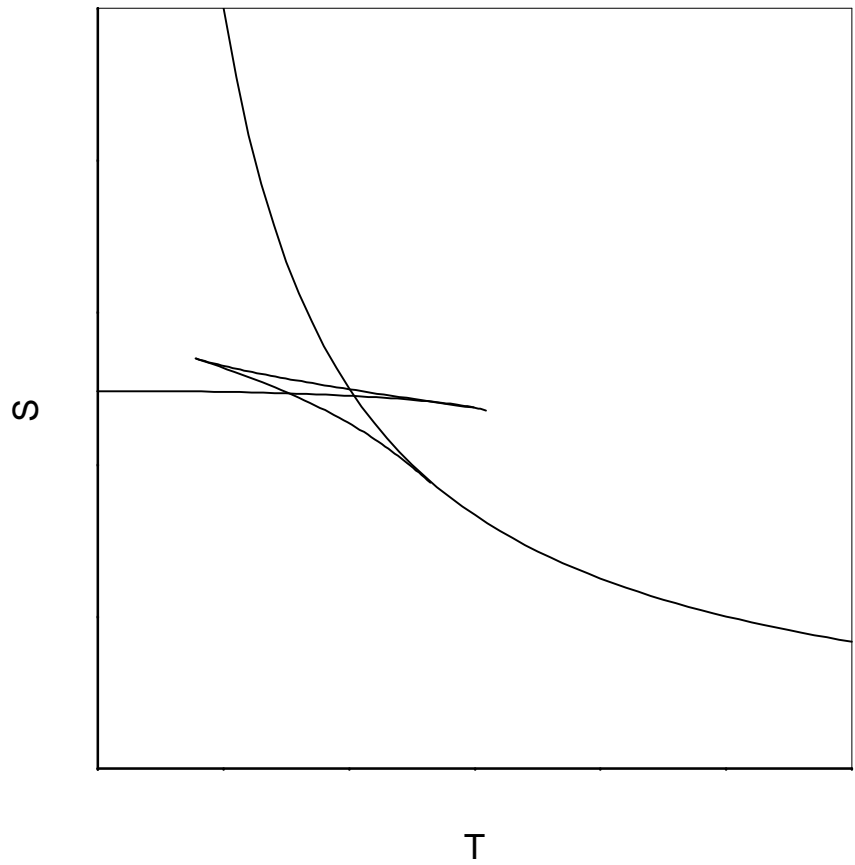
\begin{figure}
\caption{Typical three types of action-temperature diagrams. 
(a) Type I : second-order transition ( no bifurcation point ).
(b) Type II : first-order transition ( one bifurcation point ).
(c) Type III : second-order transition with an abrupt change of decay rate 
in quantum tunnelling regime ( two bifurcation points ). 
The curves showing inverse behaviour in these figures are avtion 
for sphaleron.  }
\end{figure}

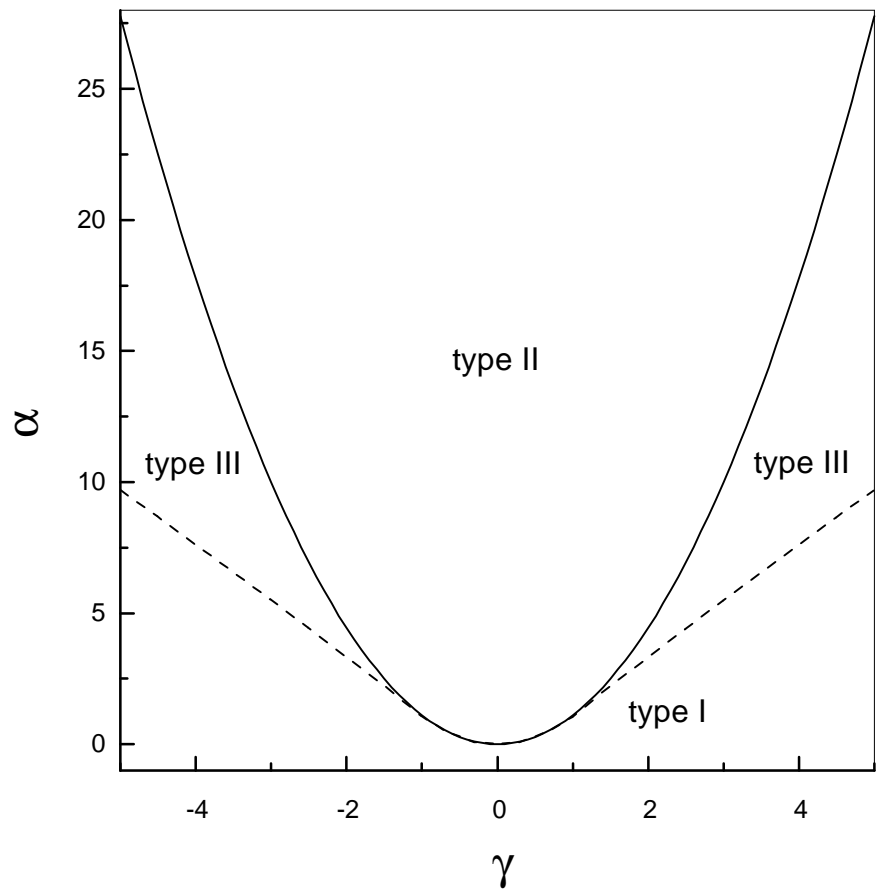
\begin{figure}
\caption{ The parameter domains of the three types for  the potential 
Eq.(\ref{qmpotential})         }
\end{figure}

\end{document}